\documentclass[]{cupconf}
\usepackage{graphicx}

\title[Metal-enrichment of HII Regions by Wolf-Rayet Stars]
      {The observable Metal-enrichment of Radiation-driven+Wind-blown 
   HII Regions in the Wolf-Rayet Stage}
      
\author[G.~Hensler et al.]
{\\ Gerhard Hensler$^1$ 
\\ Danica Kroeger$^2$
\\ Tim Freyer$^2$} 
\affiliation{$^1$ Institute of Astronomy, University of Vienna, 
A--1180 Vienna, Austria,  \\ e-mail: hensler@astro.univie.ac.at \\
 $^2$ Institute of Theor. Physics and Astrophysics, University of Kiel, 
D--24098 Kiel, Germany }   

\begin{document}
\def\HI{H{\sc i} }
\def\HII{H{\sc ii} }
\def\OIII{O[{\sc iii}] }
\def\Msun{M$_{\odot}$ }
\def\AA{A{\rm \&}A}

\maketitle

\begin{abstract}

From stellar evolution models and from observations of Wolf-Rayet stars it is known 
that massive stars are releasing metal-enriched gas in their Wolf-Rayet phase by means of
strong stellar winds. Although \HII region spectra serve as diagnostics to determine the
present-day chemical composition of the interstellar medium, it is not yet reliably 
explored to what extent the diagnostic \HII gas is already contaminated by chemically 
processed stellar wind matter. In a recent paper, we therefore analyzed our models 
of radiation-driven and wind-blown \HII bubbles around an isolated 85 \Msun star with 
originally solar metallicity with respect to its chemical abundances.
Although the hot stellar wind bubble (SWB) is enriched with $^{14}$N during the WN 
phase and even much higher with $^{12}$C and $^{16}$O during the WC phase of the star, 
we found that at the end of the stellar lifetime the mass ratios of the traced elements 
N and O in the warm ionized gas are insignificantly higher than solar, whereas an 
enrichment of 22\% above solar is found for C. The transport of enriched elements
from the hot SWB to the cool gas occurs mainly by means of mixing of hot gas with cooler 
at the backside of the SWB shell. 

\end{abstract}

\section{Introduction}
\HII regions are used as the most reliable targets to derive  
actual abundances in the Interstellar Medium (ISM). 
Kunth \& Sargent (1986) discussed the problem of determining the heavy-element
abundance of very metal-poor blue compact dwarf galaxies from emission 
lines of \HII regions in the light of local self-enrichment by massive stars
but have more stressed the effect by supernovae type II (SNeII) ejected
material. However, already during the Wolf-Rayet (WR) phase of massive
stellar evolution the stellar wind peels off the outermost stellar layers, 
so that elements from shell-burning regimes are released into the 
surrounding ISM already at later stages of their normal lifetimes. 
Since the stellar wind energetics let one presume that this gas
is deposited into the hot phase only, it was not yet reliably explored in 
detail how and to what extent the complex structure of the stellar wind 
bubble (SWB) could facilitate the cooling of wind material, by this, making
it attainable for observations of the \HII gas.. 

That WR stars should play an important role for C enrichment of the ISM
at solar metallicity was advocated by \cite{dray03} (2003). Their models 
predict that the C enrichment by WR stars is at least comparable to that by
AGB stars, while the enrichment by N is dominated by AGB stars and the
O enrichment is dominated by SNeII. Their investigation, however, sums
over all gas phases and avoids to evaluate the abundances in specific 
gas phases for detailed diagnostics, like e.g. in the warm \HII gas. 

In a series of models of radiation-driven and wind-blown bubbles produced 
by massive stars we investigated the effects of structuring and energizing 
the surrounding ISM for a 15 \Msun star (\cite{kroe07}, in preparation),
a 35 \Msun (\cite{frey06} 2006), a 60 \Msun (\cite{frey03} 2003: Paper I), 
and a 85 \Msun star (\cite{kroe06b} 2006b). From these, we could conclude that
differently strong but significant structures are formed by the combined
dynamical and radiative processes between the SWB and the enveloping
\HII region (in particular, see Paper I) where hot gas mixes with the
warm one and cools further to ''warm'' phase. The mixing occurs mainly in the
back of the SWB shell with photo-evaporated material and through turbulence
in an interface between SWB and shell (see e.g. Fig.\ref{fig1}).

Nevertheless, the WR stage is metal dependent in the sense that, 
at first, the lower the metallicity the more massive a star has to be to 
evolve through the WR stages and that,
secondly, the lower the metallicity the shorter are the WR lifetimes 
and not all WR stages are reached. 
The first point means, that the number of WR stars decreases with 
decreasing metallicity. \cite{schall92} (1992; hereafter: SSMM) found 
that for a metallicity of $Z$=0.001 the minimal zero-age main-sequence 
(ZAMS) mass for a WR star is $>$ 80 \Msun while at solar $Z$=0.02 
it is $>$ 25 \Msun as already discussed by \cite{chimae} (1986).

\section{The Model}

The hydrodynamical equations are solved together
with the transfer of H-ionizing photons on a two-dimensional cylindrical
grid. For reasons of a refined resolution mainly in the central part
around the star, the grid is structured by a nested scheme.
The time-dependent ionization and recombination of hydrogen is calculated 
in each time step and we carefully take stock of all the important energy 
exchange processes in the system. A detailed description of the
numerical method and further references are given in Paper I.

As initial condition an undisturbed homogeneous background gas with solar 
abundances (\cite{and89} 1989), hydrogen number density n$_0$=20 cm$^{-3}$, 
and temperature T$_0$=200 K is applied for the reasons described in Paper I. 
The models are then started with the sudden turn-on of the ZAMS 
stellar radiation field and stellar wind.
Since the gas is assumed as void of molecular stuff the radiation field
commences immediately to ionized the environment outwards of the SWB
without an enveloping photo-dissociation region. 
From a series of models of radiation and wind-driven \HII regions around 
single massive stars mentioned above, for our purpose the 85 \Msun star
(\cite{kroe06b} 2006b) looks as the most appropriate with respect to its 
self-enrichment. The time-dependent parameters of this star with ``standard'' 
mass-loss and solar metallicity ($Z$=0.02) during its H main-sequence and its 
subsequent evolution are taken from SSMM. 
The model analysis is already published by us (\cite{kroe06a} 2006a).

The according exploration starts not before the onset of the WR stage, in
particular with the onset of the WN stage at an age of t=2.83 Myrs. 
The WR star enriches the combined SWB/\HII region with $^{12}$C, $^{14}$N, 
and $^{16}$O. During its WN phase the star releases 0.143 \Msun $^{14}$N,
which is more than half of its total release, but nearly no extra $^{12}$C 
or $^{16}$O is supplied. As the condition for observability within the \HII 
region only the ``warm'' gas $\left(6.0 \times 10^3 \mbox{ K} \le
  T < 5.0 \times 10^4 \mbox{ K}\right)$ is accounted for. 
The mass fractions of $^{12}$C, $^{14}$N, and $^{16}$O 
with respect to solar are set according to \cite{and89} (1989) to 
$4.466 \times 10^{-3}$, $1.397 \times 10^{-3}$, and $1.061 \times 10^{-2}$, 
respectively, normalized to H.

\section{Results}

At the end of its lifetime at $t = 3.22 \mbox{ Myr}$ the 85 \Msun star has 
supplied 0.28 \Msun of $^{14}$N, 13.76 \Msun of $^{12}$C, and
11.12 \Msun of $^{16}$O, which are contained in the combined SWB/\HII 
region. Since N was released at first in the WN stage it increases slightly
in this period and is thereafter diluted by the nitrogen-poor gas feed. 
These facts are discernible in the left-hand panel of Fig. \ref{fig2} by a first 
rise and a subsequent decrease of the N abundance in the hot phase after the 
transition to the WC stage when C and O are released. The $^{12}$C
content increases steeply and reaches an overabundance of 38 times solar
while the enrichment with $^{16}$O is weaker (Fig. \ref{fig2}, left panel). 

\begin{figure}
\includegraphics[width=2.8in,height=2.1in]{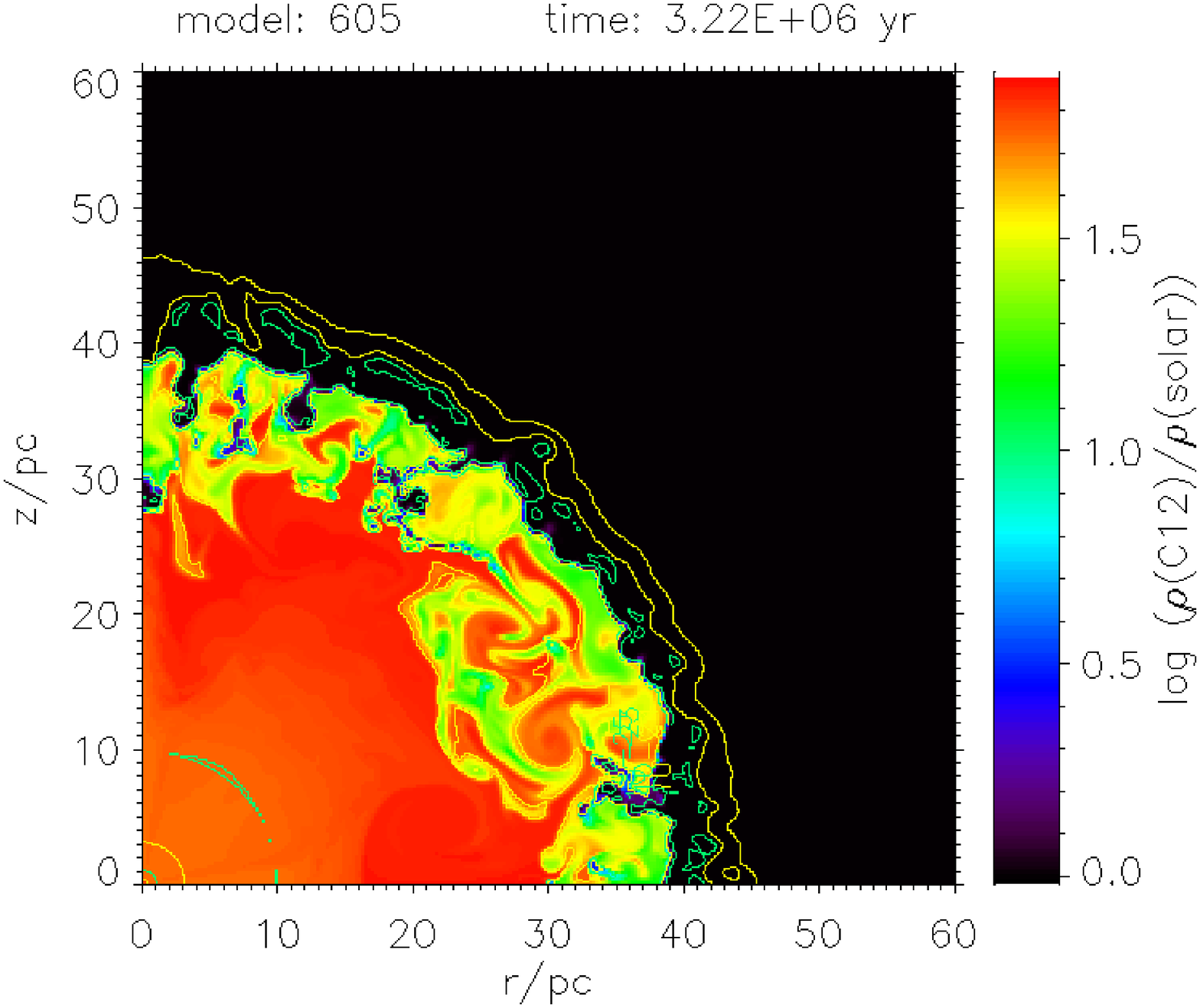}
\hfill
\includegraphics[width=2.8in,height=2.1in]{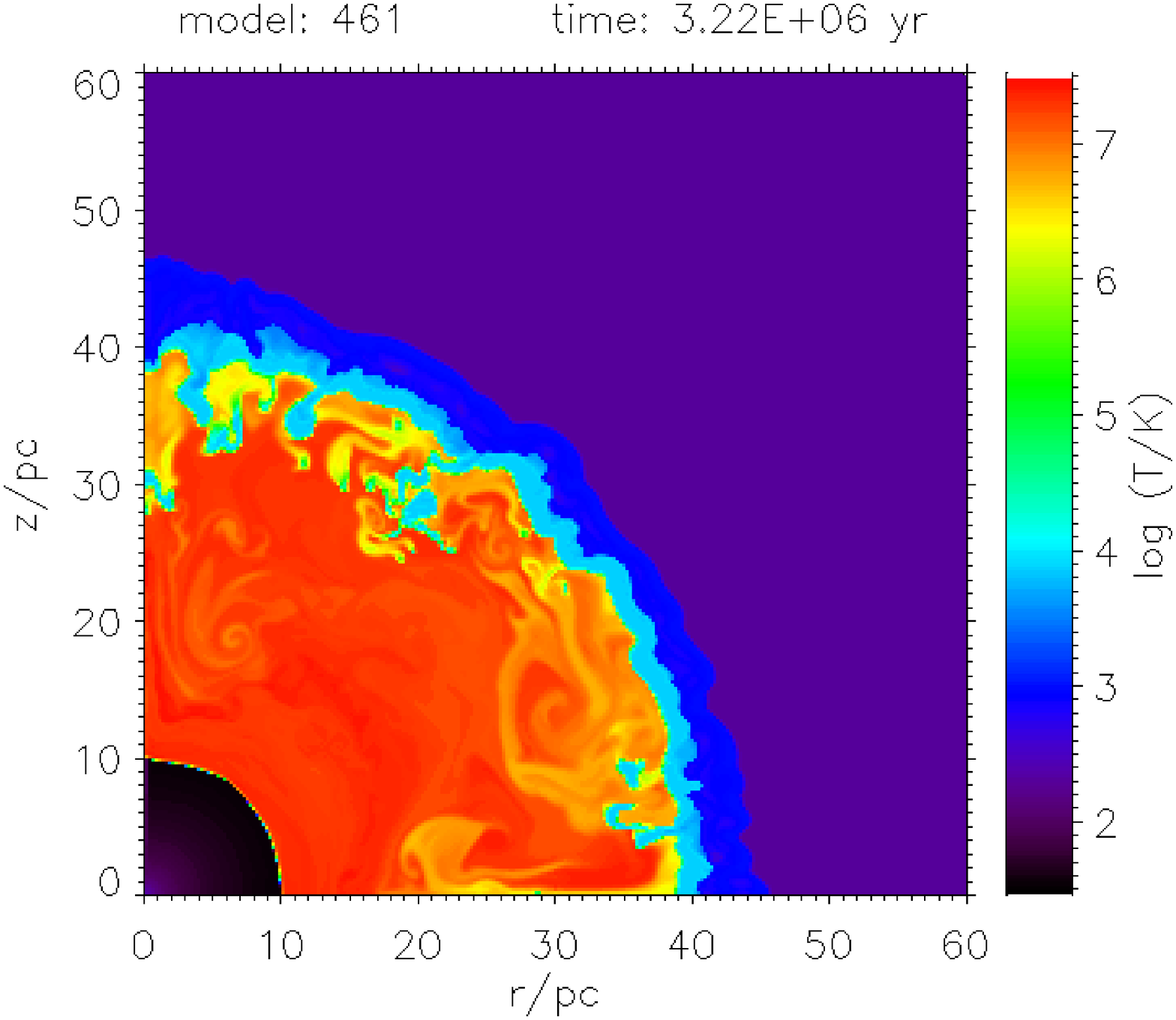}
\caption{{\it Left panel}: $^{12}$C distribution within the stellar wind bubble
and the \HII region for comparison with the temperature distribution 
({\it Right panel}). Both figures are snapshots at the end of the lifetime of a
85 \Msun\ star. All plots cover the whole computational domain of 
60 pc $\times$ 60 pc.}\label{fig1}
\end{figure}

In Fig. \ref{fig1} the C distribution as the element of largest contribution
is revealed in comparison with the temperature distribution at the end of
the stellar life. 
Two facts can be easily discerned: 1) The carbon enrichment is reasonably
largest within the hot SWB; 2) also regions with the ''warm'' temperature range
are significantly $^{12}$C enriched.

\begin{figure}
\includegraphics[width=2.2in,height=2.2in]{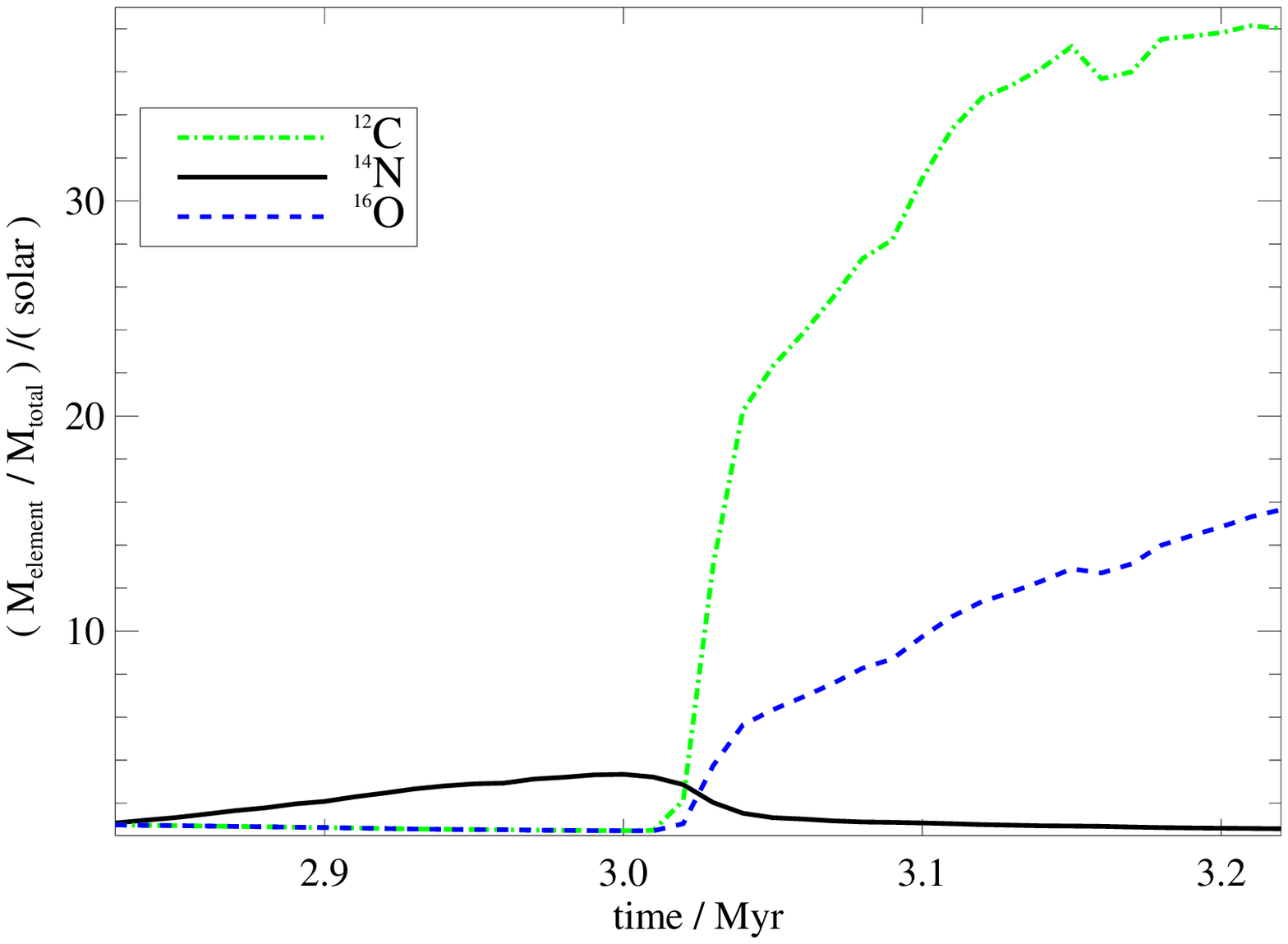}
\hfill
\includegraphics[width=2.2in,height=2.2in]{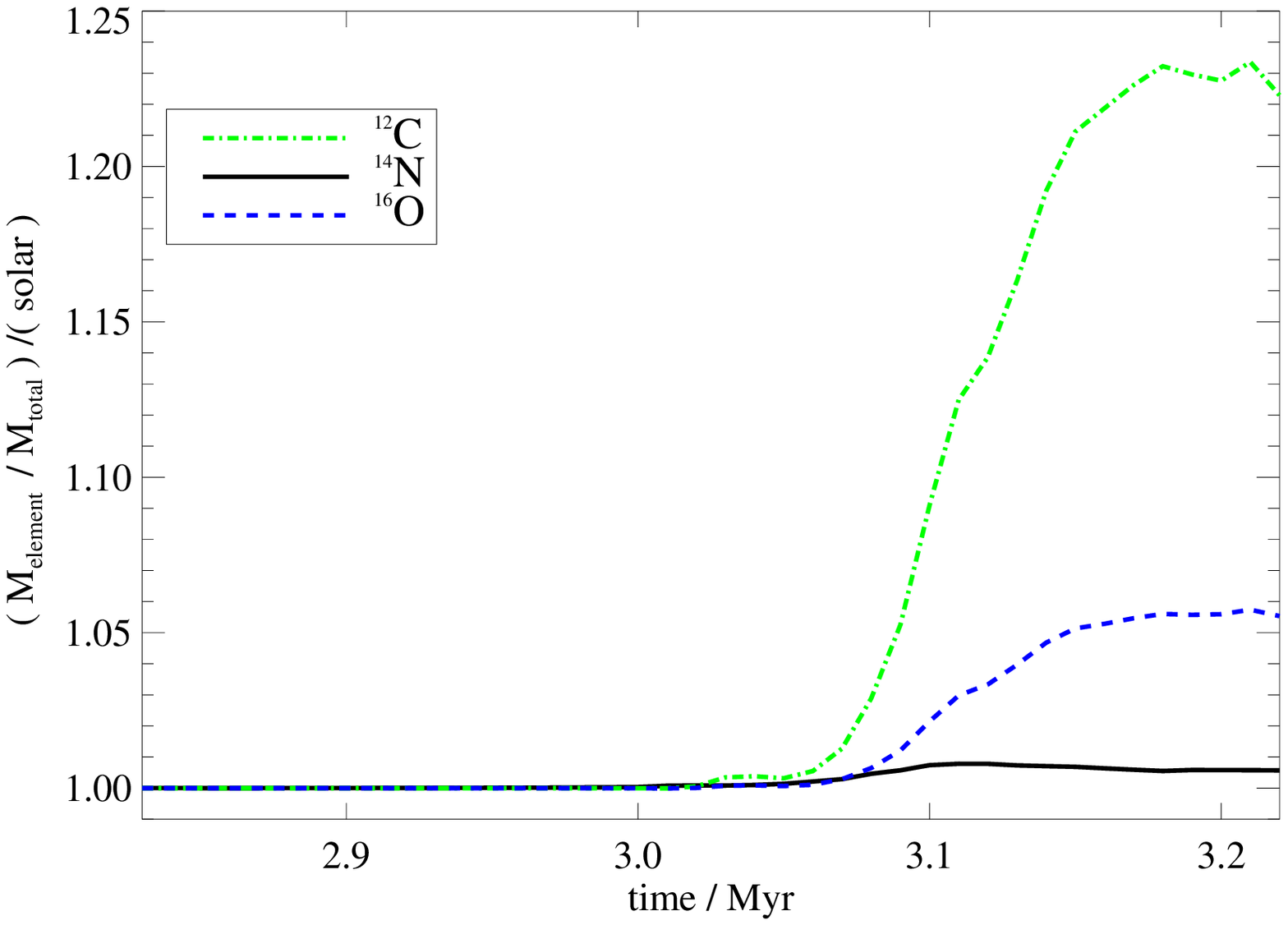}
\caption{Time-dependent abundances of $^{12}$C, $^{14}$N, and $^{16}$O 
    in the hot ({\it left panel}) and the warm gas phase ({\it right panel}). 
    The plot starts not before the onset of the WN phase at 2.83 Myrs.}\label{fig2}
\end{figure}

While the mixing and incorporation of $^{14}$N from the hot into the warm phase 
becomes only slightly detectable after about 3.1 Mrys, but occurs with a time 
delay after its release of almost 0.2 Myrs, the $^{12}$C enrichment in the warm
phase is clearly perceivible already less than 50000 yrs after its steep rise
in the hot SWB.

At the end of the 85 \Msun star's life the element quantities measurable in
emission spectra of the warm \HII region gas amount to 1.22 times solar for 
$^{12}$C, to less than 1.01 for $^{14}$N, and to only 1.05 solar for $^{16}$O.
From this model we conclude that the enrichment of the circumstellar 
environment with  $^{14}$N and  $^{16}$O by WR stars is negligible,
if the 85 \Msun star is representative for massive stars passing the WR stage.
Only for $^{12}$C the enrichment of the \HII region is significant.
For a giant \HII region containing a full set of massive stars according to
a normal initial mass function and with different lifetimes and wind mass-loss
rates the enrichment effect of C should, however, become smaller than modelled 
here for a single most massive star.
A comprehensive description and discussion of the model is already published by 
\cite{kroe06a} (2006a).

Since the occurrence of a WR phase is strongly metal dependent, the 
enrichment with C should also depend on the average metallicity. This would 
mean that any radial gradient of C abundance of \HII regions in galactic disks 
is steeper than that of O. And indeed, \cite{este05} (2005) found 
d$[\log$(C/O)]/dr = $- 0.058 \pm 0.018 \mbox{ dex kpc}^{-1}$ 
for the Galactic disk.

In metal-poor galaxies one would expect less chemical self-enrichment 
because the stellar mass range of the WR occurrence is shrinked and 
shifted rowards higher masses and the WR phases are shorter.

\section{QUESTIONS}

\noindent {\it J.M. Vilchez}: Is the $^4$He following the same behaviour
as C or O in your models?

\noindent {\it G. Hensler}: I would expect it but we haven't yet considered it.

\end{document}